\documentstyle[prl,aps,multicol,epsfig,amssymb,amsmath]{revtex}
\def\unity{\mbox{\small 1} \!\! \mbox{1}}

\begin{document}

\def\ra{\rangle}
\def\la{\langle}
\def\bege{\begin{equation}}
\def\ende{\end{equation}}
\def\begarr{\begin{eqnarray}}
\def\endarr{\end{eqnarray}}
\def\ha{{\hat a}}
\def\hb{{\hat b}}
\def\hu{{\hat u}}
\def\hv{{\hat v}}
\def\hc{{\hat c}}
\def\hd{{\hat d}}
\def\no{\noindent}\def\non{\nonumber}
\def\hi{\hangindent=45pt}
\def\v{\vskip 12pt}

\draft

\title{A Quantum Rosetta Stone for Interferometry}

\author{Hwang Lee, Pieter Kok, and Jonathan P.\ Dowling 
}

\address{
Quantum Computing Technologies Group, 
Exploration Systems Autonomy, Section 367 \\
 Jet Propulsion Laboratory,
 California Institute of Technology \\
 MS 126-347,
 4800 Oak Grove Drive, Pasadena, CA~~91109-8099
}

\date{March 31, 2002}

\maketitle

\begin{abstract}
 Heisenberg-limited measurement protocols can be used to gain an
 increase in measurement precision over classical protocols. Such
 measurements can be implemented using, e.g., optical Mach-Zehnder
 interferometers and Ramsey spectroscopes. We address the formal
 equivalence between the Mach-Zehnder interferometer, the Ramsey
 spectroscope, and a generic quantum logic circuit. Based on this
 equivalence we introduce the ``quantum Rosetta stone'', and we
 describe a projective-measurement scheme for generating the desired
 correlations between the interferometric input states in order to
 achieve Heisenberg-limited sensitivity. The Rosetta stone then tells
 us the same method should work in atom spectroscopy.
\end{abstract}

\pacs{PACS numbers: 03.65.Ud, 42.50.Dv, 03.67.-a, 42.25.Hz, 85.40.Hp}

\begin{multicols}{2}


A generic classical interferometer has a shot-noise limited
sensitivity that scales with $N^{-\frac{1}{2}}$. Here, $N$ is either
the average number of particles passing through the interferometer during
measurement time, or the number of times the experiment is performed with
single-particle inputs \cite{scully97,helstrom76}. However, when we
carefully prepare quantum correlations between the particles, the
inteferometer sensitivity can be improved by a factor of $\sqrt{N}$. That is,
the sensitivity now scales with $1/N$. This limit is imposed by the
Heisenberg uncertainty principle. For optical interferometers
operating at several milliwatts, the quantum sensitivity improvement
corresponds to an enhanced signal to noise ratio of eight orders of
magnitude. 

As early as 1981, Caves showed that feeding the secondary input port of
an optical Mach-Zehnder interferometer with squeezed light yields a
shot-noise lower than $N^{-\frac{1}{2}}$ (where $N$ is now the average
photon number) \cite{caves81}. Also, Yurke {\em et al}.\
\cite{yurke86,yurke86b} showed in 1986 that a properly correlated
Fock-state input for the Mach-Zehnder interferometer (here called the
{\em Yurke state}) could lead to a phase sensitivity of $\Delta \varphi
\simeq O(1/N)$. This improvement occurred for special values of $\varphi$. 
Sanders and Milburn, and Ou generalized this method to obtain $1/N$
sensitivity for all values of $\varphi$ \cite{sanders95,ou96}.

Subsequently, Holland and Burnett proposed the use of {\em dual Fock states} 
(of the form $|N,N\rangle$) to gain Heisenberg limited sensitivity
\cite{holland93}. This dual-Fock-state approach opened new
possibilities; in particular, Jacobson {\em et al}.\
\cite{yamamoto95}, and Bouyer and Kasevich \cite{bouyer97} showed that
the dual Fock state can also yield Heisenberg-limited sensitivity in
atom interferometry. 

A similar improvement in measurement sensitivity can be achieved in
the determination of frequency standards and spectroscopy; an atomic
clock using Ramsey's separated-oscillatory-fields technique is
formally equivalent to the optical Mach-Zhender interferometer. Here,
the two $\pi/2$-pulses represent the beam splitters. Wineland and
co-workers first showed that the best possible precision in frequency
standard is obtained by using maximally entangled states
\cite{wineland96}. Similarly, it was shown by one of us (JPD) that
this improved sensitivity can be exploited in atom-laser gyroscopes
\cite{dowling98}. 

Quantum lithography and microscopy is closely related to this enhanced
sensitivity. In practice, the bottleneck for reading and writing with
light is the resolution of the feature size, which is limited by the
wavelength of the light used. In classical optical lithography the
minimum feature size is determined by the Rayleigh diffraction limit
of $\lambda/4$, where $\lambda$ is the wavelength of the light. It has
been shown that this classical limit can be overcome by exploiting the
quantum nature of entangled photons 
\cite{yablo99,boto00,kok01,agarwal01,shih01,boyd01}. 

In classical optical microscopy too,
the finest detail that can be resolved cannot be much smaller than the
optical wavelength. using the same entangled-photons technique, it is
possible to image the features substantially smaller than the
wavelength of the light. The desired entangled quantum state for
quantum interferometric lithography yielding a resolution of
$\lambda/(4N)$ has the same form of the maximally entangled state
discussed in Ref.~11.

In this paper we present an overview of some aspects of the
enhancement by quantum entanglement in interferometeric devices, and
we describe another method for the generation of the desired quantum
states. The paper is organized as follows:

In section \ref{heisenberg}, we discuss the Heisenberg-limited
sensitivity and the standard shot-noise limit. 
Following previous work \cite{kok01a}, 
we then introduce phase estimation with quantum
entanglement. In the next section (\ref{rosetta}), we describe a
quantum Rosetta stone, based on the formal equivalence between the
Mach-Zehnder interferometer, atomic clocks, and a generic quantum logic
circuit. In section \ref{phasemeas} we discuss three different ways
of achieving the Heisenberg limit sensitivity. A brief description of
quantum interferometric lithography and the desired quantum state of
light is given in section \ref{litho}. In section \ref{projective} we
discuss quantum state preparation with projective measurements and its
application to Heisenberg-limited interferometry.

\section{
The Heisenberg uncertainty principle and parameter
  estimation}\label{heisenberg} 

In this section we briefly describe the measurement-sensitivity limits
due to Heisenberg's uncertainty principle, and how, in general,
quantum entanglement can be used to achieve this sensitivity. 
There are several stages in the procedure where quantum entanglement 
can be exploited, both in the state preparation and in the detection. 
First, we derive the Heisenberg limit, 
then we give the usual shot-noise
limit, and we conclude this section with entanglement 
enhanced parameter estimation. 

Suppose we have a $(N+1)$-level system. Furthermore, we use the
angular momentum representation to find the minimum uncertainty
$\Delta Q$ in an observable ${\hat Q}$ that is a dual to the angular momentum
operator ${\hat J}_z$. That is, ${\hat Q}$ and ${\hat J}_z$ 
obey a Heisenberg uncertainty
relation ($\hbar =1)$:
  
\begin{equation}\label{hur}
  \Delta Q\, \Delta J_z \geq \frac{1}{2}\; .
\end{equation}

\no
When we want minimum uncertainty in ${\hat Q}$ (minimal $\Delta Q$), 
we need to maximize the uncertainty in 
${\hat J}_z$ (maximal $\Delta {J}_z$). 
Given
the eigenstates $\{|m\rangle\}_{m=-N/2}^{+N/2}$ of ${\hat J}_z$: 
${\hat J}_z |m\rangle = m |m\rangle$, 
let us, for example, take the state of the form
$|\psi\rangle = (N+1)^{-\frac{1}{2}}\sum_{m=-N/2}^{+N/2} e^{i\phi_m}
|m\rangle$, which consists of
equally distributed eigenstates of ${\hat J}_z$. 
The variance in ${\hat J}_z$ is then given by 

\begin{equation}
  (\Delta {J}_z)^2 = 
\langle\psi|{\hat J}_z^2|\psi\rangle -
  \langle\psi|{\hat J}_z|\psi\rangle^2 = \frac{1}{3}
\left( {N^2 \over 4} + {N \over 2} \right)\; .
\end{equation}
It immediately follows that the leading term in $\Delta J_z$ scales
with $N$.

If we choose the quantum state
$|\psi\rangle = 2^{-\frac{1}{2}}
({|m={N \over 2}\rangle} + e^{i\phi}|{m=-{N\over 2}\rangle})$, 
we obtain

\begin{equation}
  (\Delta J_z)^2 =
 {N^2 \over 4} \; .
\end{equation}

\no
Using the equality sign in Eq.\ (\ref{hur}), i.e., minimum uncertainty,
and the expression for $\Delta J_z$, we find that 
$\Delta Q = 1/N$.

By contrast consider a quantum state of the form
$|\psi\rangle = 2^{-\frac{N}{2}}\sum_{m=-N/2}^{+N/2} e^{i\phi_m}
(_N C_{m+N/2})^{1 \over 2} |m\rangle$, which is a 
binomial distribution of the eigenstates of ${\hat J}_z$. 
The variance in ${\hat J}_z$ in this case is given by 

\begin{equation}
  (\Delta J_z)^2 =
 {N \over 4} \; ,
\end{equation}

\no
which, then, corresponds to the usual shot-noise limit.

This result gives the spread of measurement outcomes of an observable
${\hat Q}$ in a $(N+1)$-level system. However, it is not yet cast in the
language of standard parameter estimation. The next question is
therefore how to achieve this {\em Heisenberg-limited} sensitivity
when we wish to estimate the value of a parameter $\varphi$ in $N$
trials.  

The shot-noise limit (SL), according to estimation theory, 
is given by $\Delta\varphi=N^{-\frac{1}{2}}$. 
We give a short derivation of this value and generalize it 
to the quantum mechanical case. 
Consider an ensemble of $N$ two-level systems in the state
$|\varphi\rangle = (|0\rangle+e^{i\varphi}|1\rangle)/\sqrt{2}$, where
$|0\rangle$ and $|1\rangle$ denote the two levels. If $\hat{A} =
|0\rangle\langle 1| + |1\rangle\langle 0|$, then the expectation value
of $\hat{A}$ is given by

\begin{equation}
  \langle\varphi|\hat{A}|\varphi\rangle=\cos\varphi\; .
\end{equation}

\no
When we repeat this experiment $N$ times, we obtain
\begin{equation}
  _N\langle\varphi|\ldots\, _1\!\langle\varphi| \left(
  \,\overset{N}{\underset{k=1}{\mbox{\Large $\oplus$}}}\,
  \hat{A}^{(k)}\right)|\varphi\rangle_1 \ldots |\varphi\rangle_N = N
  \cos\varphi\; .
\end{equation} 

\no
Furthermore, it follows from the definition of $\hat{A}$ that
$\hat{A}^2=\unity$ on the relevant subspace, and
the variance of $\hat{A}$ given $N$ samples is readily computed to be
$(\Delta A)^2 = N(1-\cos^2\varphi) = N \sin^2 \varphi$. According to
estimation theory \cite{helstrom76}, we have 

\begin{equation}\label{est}
  \Delta\varphi_{SL} = \frac{\Delta A}{|d\langle
  \hat{A}\rangle/d\varphi|} = \frac{1}{\sqrt{N}}\; .
\end{equation}

\no
This is the standard variance in the parameter $\varphi$ after $N$
trials. In other words, the uncertainty in the phase is inversely
proportional to the square root of the number of trials. 
This is, then, the shot-noise limit.

Quantum entanglement can improve the sensitivity of this procedure by a
factor of $\sqrt{N}$. We will first employ an entangled state

\begin{equation}\label{entang}
 |\varphi_N\rangle\equiv|N,0\rangle + e^{iN\varphi}|0,N\rangle\; , 
\end{equation}

\no
where $|N,0\rangle$ and $|0,N\rangle$ are 
collective states of $N$ qubits, defined,
for example, 
in the computational basis $\{|0\ra$, $|1\ra\}$, as

\begarr
|N,0\ra &=& |0\ra_1 |0\ra_2 \cdot\cdot\cdot |0\ra_N \non \\
|0,N\ra &=& |1\ra_1 |1\ra_2 \cdot\cdot\cdot |1\ra_N .
\label{n0}
\endarr

\no 
The relative phase $e^{iN\varphi}$ can be obtained 
by a unitary evolution of one of the modes of $|\varphi_N\rangle$.
Here, in the case of Eq.\ (\ref{entang}),
each qubit in state $|1\ra$ acquires a phase shift of $e^{i \varphi}$. 
When we measure the observable
$\hat{A}_N = |0,N\rangle\langle N,0| + |N,0\rangle\langle 0,N|$ we have
  
\begin{equation}\label{cosn}
  \langle\varphi_N |\hat{A}_N| \varphi_N\rangle = \cos N\varphi\; .  
\end{equation}

\no
Again, $\hat{A}_N^2=\unity$ on the relevant subspace, and 
$(\Delta A_N)^2 = 1-\cos^2 N\varphi = \sin^2 N\varphi$.
Using Eq.\ (\ref{est}) again, we obtain the so-called 
Heisenberg limit (HL) to the minimal detectable phase: 

\begin{equation}\label{bol}
  \Delta\varphi_{HL} = \frac{\Delta A_N}{|d\langle \hat{A}_N
  \rangle/d\varphi|}=\frac{1}{N}\; .
\end{equation}

\no
The precision in $\varphi$ is increased by a factor $\sqrt{N}$ over
the standard noise limit.
Detailed descriptions of phase estimation have been given 
by Hradil \cite{hradil92}, and Lane, Braunstein, and Caves \cite{lane93}. 

As shown in Bollinger {\em et al}.\
\cite{wineland96}, Eq.\ (\ref{bol}) is the optimal accuracy permitted
by the Heisenberg uncertainty principle. 
In quantum lithography, one exploits the $\cos(N\varphi)$ behaviour, 
exhibited by Eq.\ (\ref{cosn}), 
to print closely spaced lines on a suitable substrate \cite{boto00}. 
Gyroscopy and entanglement-enhanced frequency
measurements \cite{dowling98} exploit the $\sqrt{N}$ increased
precision. The physical interpretations of $A_N$ and the phase
$\varphi$ will differ in the different protocols. In the following
section we present three distinct physical representations of
this construction. We call this the {\em quantum Rosetta stone}. Of
particular importance is the interplay between the created states
$|\varphi_N\rangle$ and the measured observable $\hat{A}_N$ in
Eq.~(\ref{cosn}). In section \ref{phasemeas} we present three types of
quantum states and observables that yield Heisenberg-limited
sensitivity.

\section{
Quantum Rosetta stone}\label{rosetta}

In this section we discuss the formal equivalence between the
Mach-Zehnder interferometer, the Ramsey spectroscope, and a logical
quantum gate.

\begin{figure}[htb]
\epsfysize=8cm
\centerline{\epsfbox{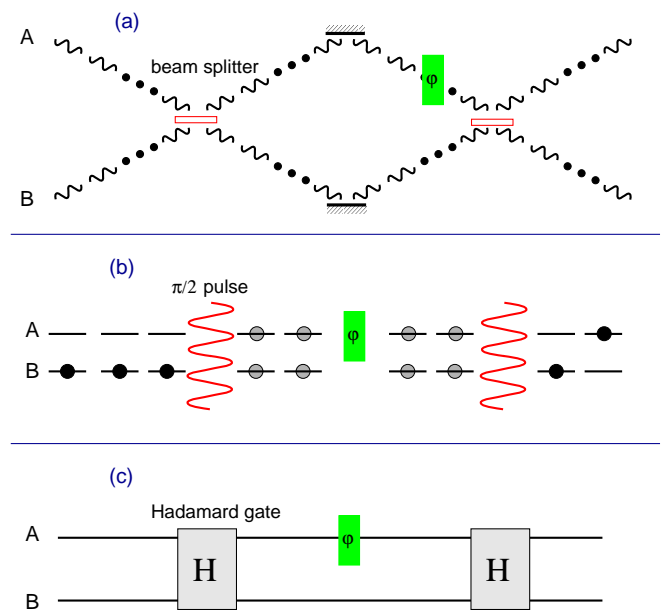}}

\v
\caption{\label{fig-1}
The quantum Rosetta stone.
(a) An optical Mach-Zehnder interferometer.
(b) Ramsey atomic clock.
(c) A quantum logic gate representing
the equivalent physical proccess.}
\end{figure}

In a Mach-Zehnder interferometer \cite{note}, the input light field is divided
into two different paths by a beam splitter, and recombined by another
beam splitter. The phase difference between the two paths is then
measured by balanced detection of the two output modes (see
Fig.~\ref{fig-1}a). 
A similar situation, which we will omit in our discussion,
can be found in Stern-Gerlach filters
in series \cite{feynman79}, and
the technical limitations of
such a device has been
discussed by Englert, Schwinger, and Scully \cite{ess88}.

With a coherent laser field as the input the phase
sensitivity is given by the shot noise limit $N^{-\frac{1}{2}}$, where
$N$ is the average number of photons passing though the interferometer
during measurement time. When the number of photons is exactly known
(i.e., the input is a Fock state $|N\rangle$), the phase sensitivity
is still given by $N^{-\frac{1}{2}}$, indicating that the photon
counting noise does not originate from the intensity fluctuations of
the input beam \cite{dowling98,scully93}. In the next section we will
see how we can improve this sensitivity.

In a Ramsey spectroscope, atoms are put in a superposition of the
ground state and an excited state with a $\pi/2$-pulse
(Fig.~\ref{fig-1}b). After a time interval of free evolution, a second
$\pi/2$-pulse is applied to the atom, and, depending on the relative
phase shift obtained by the excited state in the free evolution, the
outgoing atom is measured either in the ground or the excited
state. Repeating this procedure $N$ times, the phase $\varphi$ is
determined with precision $N^{-\frac{1}{2}}$. This is essentially an atomic 
clock. Both procedures are methods to measure the phase shift, either
due to the path difference in the interferometer, or to the
free-evolution time in the spectroscope. When we use entangled atoms
in the spectroscope, we can again increase the sensitivity of the
apparatus.

A third system is given by a qubit that undergoes a Hadamard transform
$H$, then picks up a relative phase and is then transformed back with
a second Hadamard transformation (Fig.~\ref{fig-1}c). This
representation is more mathematical than the previous two, and it
allows us to extract the unifying mathematical principle that connects
the three systems. In all protocols, the initial state is transformed
by a discrete Fourier transform (beam splitter, $\pi/2$-pulse or
Hadamard), then picks up a relative phase, and is transformed back
again. 
This is the standard ``quantum'' finite Fourier transform, such as used 
in the implementation of Shor's algorithm \cite{ekert96}. 
It is not the same as the 
classical algorithm in engineering.

As a consequence, the phase shift (which is hard to measure
directly) is applied to the transformed basis. The result is a bit flip
in the initial, {\em computational}, basis $\{ |0\rangle,|1\rangle\}$,
and this is readily measured. We call the formal analogy between these
three systems the {\em quantum Rosetta stone}.

The importance of the Rosetta stone was that, by giving an example of 
writing in three different languages: 
Greek, Demotic, and Hieroglyphics, 
it enabled the
French scholar Jean Fran\c{c}ois Champollion to crack the code 
of the hieroglyphs,
which was not understood until his work in 1822 \cite{champollion22}. 
In discussing quantum
computer circuits with researchers from the fields of quantum optics or atomic
clocks, we find the ``quantum Rosetta stone'' a useful tool to connect those areas
of research to the language of quantum computing -- which can be as
indecipherable as hieroglyphs to such researchers.

These schemes can be generalized from measuring a simple phase shift to
evaluating the action of a unitary transformation $U_f$ associated
with a complicated function $f$ on multiple qubits. 
Such an evaluation is also
known as a quantum computation. 
Quantum computing with generalized 
Mach-Zehnder interferometers has been proposed 
by Milburn \cite{milburn89}, and later
linear optics quantum computation has been suggested
by Cerf, Adami, and Kwiat \cite{cerf98}.
Recently, Knill, Laflamme and Milburn
demonstrated an efficient linear optics 
quantum computation \cite{milburn01}. 
Implementations of quantum logic gates have
also been proposed by using atoms and ion traps
\cite{haroche95,cirac95,kimble95,monroe95}.
According to our Rosetta stone, the concept 
of quantum computers is therefore to exploit quantum interference in
obtaining the outcome of a computation of $f$. In this light, a
quantum computer is nothing but a (complicated) multiparticle quantum
interferometer \cite{ekert98,cleve98}.

This logic may also be reversed:
a quantum interferometer is therefore a (simple) quantum computer.
Often, when discussing, Heisenberg-limited interferometry
with complicated entangled states,
one encounters the critique that such states are
highly susceptible to noise and even one or
a few uncontrolled interactions with the environment
will cause sufficient degradation of the device and
recover only the shot-noise limit \cite{kimble00}.
Apply the {\em quantum Rosetta stone} by replacing the term
``quantum interferometer'' with ``quantum computer''
and we recognize the exact same critique that
has been leveled against quantum computers for years.
However, for quantum computers, we know the response--to apply
quantum error-correcting techniques and encode in decoherence
free subspaces.
Quantum interferometry is just as hard as quantum computing!
The same error-correcting tools that we believe
will make quantum computing a reality,
will also be enabling for quantum interferometry.

\section{
Quantum enhancement in phase measurements}\label{phasemeas}

There have been various proposals for achieving Heisenberg-limited
sensitivity, corresponding to different physical realizations of the
state $|\varphi_N\rangle$ and observable $\hat{A}_N$ in Eq.~(\ref{cosn}).
Here, we discuss three different approaches, categorized according to
the different quantum states. We distinguish Yurke states, dual Fock
states, and maximally entangled states.

\subsection{Yurke states}\label{yurke-sec}

By utilizing the $su(2)$ algebra of spin angular momentum, 
Yurke {\em  et al}.\ have shown that, 
with a suitably correlated input state, 
the phase sensitivity can be improved to $1/N$ \cite{yurke86,yurke86b}. 
Similarly, Hillery and Mlodinow,
and Brif and Mann
have proposed using the
so-called minimum uncertainty state
or the ``intelligent state''.
The minimum uncertainty state is defined as
$\Delta J_x \Delta J_y = |\la {\hat J}_z \ra |/2$,
and such a state with $\Delta J_y \rightarrow 0$
can yield the Heisenberg limited sensitivity
under certain conditions \cite{hillery93,brif96}.

Let ${\hat a}^\dagger$, ${\hat b}^\dagger$ denote the 
creation operators for the two input modes in Fig.~\ref{fig-1}a. 
In the Schwinger representation, 
the common eigenstates of ${\hat J}^2$ and ${\hat J}_z$ 
are the two-mode Fock states $|j,m\ra = |j+m\ra_A |j-m\ra_B$,
where 
\begarr
 {\hat J}_x & = &( {\hat a}^\dagger {\hat b} + {\hat b} {\hat
   a}^\dagger )/2\; ; \quad {\hat J}_y = -i( {\hat a}^\dagger {\hat b} -
 {\hat b} {\hat a}^\dagger )/2\; ; \non \\
 {\hat J}_z & = &( {\hat a}^\dagger {\hat a} - {\hat b}^\dagger {\hat
   b} )/2\; ; \quad {\hat J}^2 = {\hat J}_x^2 + {\hat J}_y^2 + {\hat
  J}_z^2\; .
\label{jm}
\endarr
The interferometer can be described by the rotation of the angular
momemtum vector, where ${\hat a}^\dagger {\hat a} + {\hat b}^\dagger 
{\hat b}= N = 2j$, and the 50/50 beam splitters and the phaser shift
are corresponding to the operators $e^{i \pi {\hat J}_x/2}$ and
$e^{i\varphi {\hat J}_z}$, respectively.  

For spin-1/2 fermions, the entangled input state (which we call the
`Yurke state') $|\varphi_N\ra_{\rm Y}$ is given by
\begarr
 |\varphi_N\ra_{\rm Y} &=& {1 \over \sqrt{2}} \left[ \left|j=\mbox{$\frac{N}{2}$}, m
     =\mbox{$\frac{1}{2}$}\right\ra + \left|j=\mbox{$\frac{N}{2}$},
     m=-\mbox{$\frac{1}{2}$} \right\ra \right] \non \\ 
 &=& {1 \over \sqrt{2}} \left[
   \left|\mbox{$\frac{N+1}{2}$},\mbox{$\frac{N-1}{2}$} 
   \right\ra_{AB} +
   \left|\mbox{$\frac{N-1}{2}$},\mbox{$\frac{N+1}{2}$} \right\ra_{AB} 
 \right],
\label{yurke}
\endarr

\no
where the notion of $|j,m\ra$ follows the definition given in
Eq.~(\ref{jm}) and the subscripts $AB$ denote the two input modes. For
bosons, a similar input state, namely $|j=N/2,m=0\ra + |j=N/2,m=1\ra$,
has been proposed \cite{yurke86b}. The measured observable $\hat{A}_N$
is given by ${\hat J}_z$. After evolving the state $|\varphi_N\ra_{\rm Y}$
(in the Ramsey spectroscope or the Mach-Zehnder interferometer with
phase shift $\varphi$), the phase sensitivity $\Delta\varphi$ can be
determined to be proportional to $1/N$ for special values of
$\varphi$ (Fig.\ 2a). 
Although the input state of Eq.~(\ref{yurke}) was proposed
for spin-1/2 fermions, the same state with bosons also yields the
phase sensitivity of the order of $1/N$ \cite{dowling98}.

\subsection{Dual Fock states}

In order to achieve Heisenberg-limited sensitivity, Holland and
Burnett proposed the use of so-called {\em dual Fock states}
$|N\rangle_A \otimes |N\rangle_B$ for two input modes $A$
and $B$ of the Mach-Zehnder interferometer \cite{holland93}. Such a
state can be generated, for example, by degenerate parametric down
conversion, or by optical parametric oscillation \cite{kim98}. 

To obtain increased sensitivity with dual Fock states, 
some special detection scheme is needed.
In a conventional Mach-Zehnder
interferometer only the difference of the number of photons at the
output is measured. 
Similarly, in atom interferometers, measurements are performed by
counting the number of atoms in a specific internal state using
fluorescence.
For the schemes using dual Fock-state input, an
additional measurement is required since the average in the intensity
difference of the two output ports does not contain information about
the phase shift (Fig.\ 2b).

\begin{figure}[htb]
\epsfysize=9cm
\centerline{\epsfbox{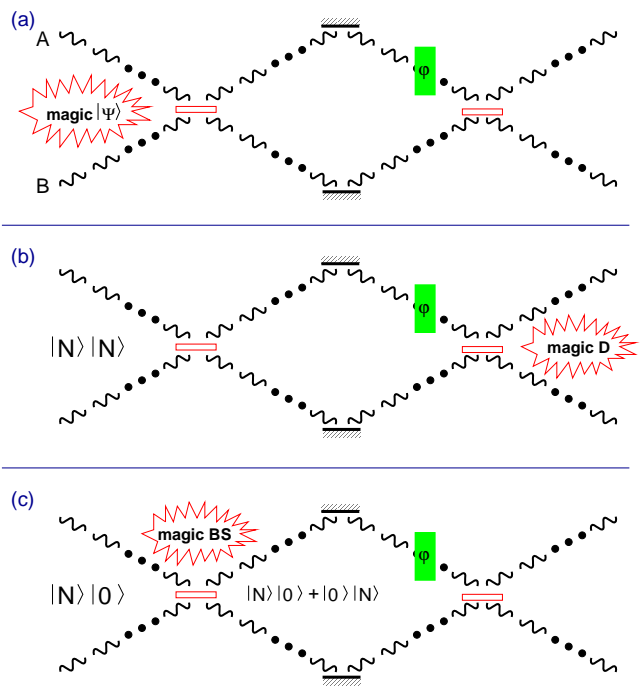}}
\bigskip
\caption{\label{fig-2}
Three catagories for achieving the Heisenberg limited
phase sensitivity.
Emphasis on the distinctive features are termed
as ``magic''.
(a) Correlated input state (magic state),
(b) dual Fock-state input (magic detectors),
(c) maximally correlated state (magic beam splitter).
}
\end{figure}

One measures both the sum and the difference
of the photon number in the two output modes \cite{holland93}. 
The sum contains information about the total photon number, 
and the difference contains
information about the phase shift. Information about the total photon
number then allows for post-processing the information about the
photon-number difference.
A combination of a direct measurement of the variance
of the difference current and a data-processing method based on
Bayesian analysis was proposed by Kim {\em et al}.\ \cite{kim98}.
Recently, Wisemen and co-workers have proposed 
an adaptive measurement scheme, where the optimal input state, however,
is different from the dual Fock states \cite{berry00}.

For atom interferometers, a quantum nondemolition measurement is
required to give the total number of atoms \cite{bouyer97}. In a
similar context, Yamamoto and co-workers devised an atom interferometry
scheme that uses a squeezed $\pi/2$ pulse for the readout of the input state
correlation \cite{yamamoto95}. 

Due to its simple form, the dual Fock-state approach may shed a new
light on Heisenberg-limited interferometry. In particular, exploiting
the fact that atoms in a Bose-Einstein condensate can be represented
by Fock states, Bouyer and Kasevich, as well as Dowling, 
have shown that the quantum noise
in atom interferometry using dual Bose-Einstein consensates can also
be reduced to the Heisenberg limit \cite{bouyer97,dowling98}.

\subsection{Maximally entangled states}

The third, and last, category of states is given by the maximally
entangled states. There have been proposals to overcome the
standard shot noise limit by using spin-squeezed states
\cite{kitagawa91,wineland92,kitagawa93,itano93,agarwal94}. However, it
was demonstrated by Wineland and co-workers \cite{wineland96}
that the optimal
frequency measurement can be achieved by using {\it maximally
entangled states} \cite{mermin90}, 
which have the following form: 
\begin{equation}\label{wineland}
 |\varphi_N\rangle = \frac{1}{\sqrt{2}}\left( \left|N,0 \right\ra_{AB}
   + \left|0,N \right\ra_{AB}\right)\; .
\end{equation}
This state has an immediate resemblence with the state in
Eq.~(\ref{cosn})
after acquiring a phase shift of $e^{iN\varphi}$. 
Although partially entangled states with a high
degree of symmetry were later shown to have a better resolution in
the presence of decoherence \cite{huelga97}, these maximally
correlated states are of great interest for optical interferometry.

\begin{figure}[htb]
\epsfysize=3.5cm
\centerline{\epsfbox{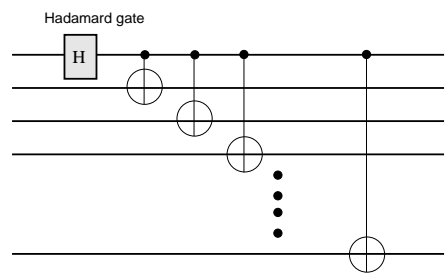}}
\bigskip
\caption{\label{fig-3}
Quantum logic gates for generation of
the maximally entangled states (representing
the ``magic'' beam splitter in Fig. 2c).
$N$ qubits become maximally entangled via
a Hadamard gate followed by $N$ C-NOT gates.
}
\end{figure}

In terms of quantum logic gates, the maximally correlated state of the
form of Eq.~(\ref{wineland}) can be made using a Hadamard gate and a
sequence of C-NOT gates (see Fig.\ 3). 
Note that
one distinctive feature, compared to the other schemes described above,
is that the state of the form Eq.\ (\ref{wineland})
is the desired quantum state after the first beam splitter
in the Mach-Zehnder interferometer, not the input state (Fig.\ 2c). 
In that the
desired input state is described as the inverse beam-splitter
operation to the state of Eq.~(\ref{wineland}).

All the interferometric schemes using entangled or dual-Fock input
states show a sensitivity approaching $1/N$ only asymptotically. 
However, using the maximally correlated states of Eq.~(\ref{wineland}),
the phase sensitivity is equal to $1/N$, even for a small $N$.

\section{Quantum lithography and maximally
path-entangled states}\label{litho}

Quantum correlations can also be applied to optical lithography. 
In recent work it has been shown that the Rayleigh diffraction limit
in optical lithography can be circumvented by the use of
path-entangled photon number states \cite{yablo99,boto00}. The desired
$N$-photon path-entangled state, for $N$-fold resolution enhancement,
is again of the form\footnote{
We call the state of the form  $|N,0\ra +|0,N\ra$ 
as the NOON state,
and the High NOON state for a large $N$.}
 given in Eqs.~(\ref{entang}) and (\ref{wineland}).

Consider the simple case of a two-photon Fock state $|1\ra_A |1\ra_B$,
which is a natural component of a spontaneous parametric
down-conversion event. After passing through a 50/50 beam splitter,
it becomes an entangled number state of the form $|2\ra_A |0\ra_B +
|0\ra_A |2\ra_B$. Interference suppresses the probability amplitude 
of $|1\ra_A |1\ra_B$. According to quantum mechanics, it is not
possible to tell whether both photons took path $A$ or $B$ after the
beam splitter. 

When parametrizing the position $x$ on the surface by $\varphi=\pi
x/\lambda$, the deposition rate of the two photons onto the substrate
becomes $1+\cos 2\varphi$, which has twice better resolution
$\lambda/8$ than that of single-photon absorption, $1 + \cos\varphi$,
or that of uncorrelated two-photon absorption, $(1 +\cos\varphi)^2$.
For $N$-photon path-entangled state of Eq.~(\ref{wineland}), we obtain
the deposition rate $1+\cos N\varphi$, corresponding to a resolution
enhancement of $\lambda/(4N)$. 

It is well known that the two-photon path-entangled state of
Eq.~(\ref{wineland}) can be generated using a Hong-Ou-Mandel (HOM)
interferometer \cite{mandel87} and two single-photon input states.
A 50/50 beam splitter, however, is not sufficient for producing
path-entangled states with a photon number larger than two
\cite{campos89}. 

As depicted in Fig.\ 3, the maximally correlated state of the
form of Eq.~(\ref{wineland}) can be made using a Hadamard gate and a
sequence of $N$ C-NOT gates. However, building 
optical C-NOT gates normally
requires  large optical nonlinearities. 
Consequently, in generating such states
it is commonly assumed that large $\chi^{(3)}$ nonlinear 
optical components are needed for $N >2$.

Knill, Laflamme, and Milburn proposed a method for creating
probabilistic single-photon quantum logic gates based on
teleportation. The only resources for this method are linear optics
and projective measurements \cite{milburn01}. Probabilistic quantum
logic gates using polarization degrees of freedom have been proposed by
Imoto and co-workers, and Franson's team \cite{imoto01,franson01}. 
In particular, Pittman, Jacobs, and Franson have experimentally
demonstrated polarization-based C-NOT implementations \cite{franson02}.

Using the concept of projective measurements, we have previously
demonstrated that the desired
path-entangled states can be created
when conditioned on the measurement outcome \cite{kok01a,lee01}.
This way, one can avoid the use of large $\chi^{(3)}$ nonlinearities
\cite{gerry01}.
In the next section we discuss the path-entanglement generation
based on projective measurements, and
its application to Heisenberg-limited interferometry.

\section{Projective measurements and Heisenberg-limited
  interferometry}\label{projective} 

As is discussed in the previous sections,
the Yurke state approach has the same measurement scheme as
the conventional Mach-Zehnder interferometer
(direct detection of the difference current), but
it is not easy to generate the desired correlation
in the input state \cite{dowling98}.
On the other hand, the dual Fock-state
approach finds a rather simple input state, but
requires a complicated measuremet schemes.
In this section we describe a method for creating a desired
correlation in the Yurke state approach 
directly from the dual Fock state
by using the projective measurements.

\begin{figure}[htb]
\epsfysize=5cm
\centerline{\epsfbox{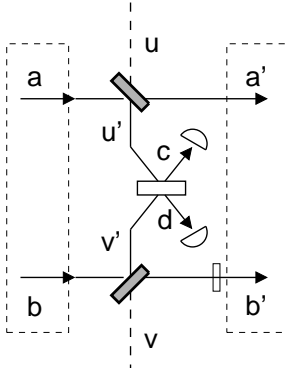}}
\caption{\label{gizmo} 
If one and only one photon is detected at each detector,
two photons are ``peeled off'' from either input mode $a$ or $b$.
}
\end{figure}

Consider the scheme depicted in Fig.~\ref{gizmo}. The input modes are
transformed by the beam splitters as follows:
\begarr  
 {\hat a}^\dagger &&\rightarrow i t {\hat a}^{\prime \dagger} + r
 {\hat u}^{\prime\dagger}, \non \\
 {\hat b}^\dagger && \rightarrow i t {\hat b}^{\prime\dagger} + r
 {\hat v}^{\prime\dagger}, 
\endarr
where $i=\sqrt{-1}$, and $it$ and $r$ are the transmission and
reflection coefficients given by $t^2+r^2=1$. In our convention, a
50/50 beam splitter, for example, is identified as $t=1/\sqrt{2}$ and
$r=-1/\sqrt{2}$.

The modes $\hu^\prime$ and $\hv^\prime$ further pass
through an additional 50/50 beam splitter, which is characterized
by the transformations:
\begarr  
{\hat u}^{\prime\dagger} &&\rightarrow
 (i  {\hat d}^{\dagger}
-  {\hat c}^{\dagger} )/\sqrt{2}, \non\\
{\hat v}^{\prime\dagger} && \rightarrow
 (i  {\hat c}^{\dagger} 
- {\hat d}^{\dagger} )/\sqrt{2}.
\label{uv}
\endarr

Suppose the input state is given by the dual Fock state
\begarr
|\Psi_{\rm in}\ra = |N,N\ra =
{ (\ha^\dagger)^N (\hb^\dagger)^N \over N!} |0\ra \; . 
\endarr

\no
Accordingly, the beam splitter transformation, e.g., for the mode $a$,
can be written as

\begarr
(\ha^\dagger)^N 
&\rightarrow&
\sum_k ~_N C_k (\ha^{\prime\dagger})^{N-k} 
(\hu^{\prime\dagger})^k r^k (it)^{N-k}\; .
\label{an}
\endarr

\no
Let both modes pass through a beam splitter, and recombine the
reflected modes $u'$ and $v'$ in a 50/50 beam splitter as depicted
in Fig.\ \ref{gizmo}. 
We post-select
events on a two-fold detector coincidence at the output modes of this
beam splitter. Assuming ideal detectors, this will restrict the final
state of the outgoing modes $a'$ and $b'$. Then, from Eq.~(\ref{uv}),
we have 
\begarr  
 {\hat u}^{\prime\dagger} {\hat v}^{\prime\dagger} &&\rightarrow
 -i \left[ (\hc^\dagger)^2 + (\hd^\dagger)^2 \right]/2\; , \non \\
 ({\hat u}^{\prime\dagger})^2 && \rightarrow \left[ (\hc^\dagger)^2  -
   2i \hc^\dagger \hd^\dagger - (\hd^\dagger)^2 \right]/2\; , \non \\
 ({\hat v}^{\prime\dagger})^2 && \rightarrow \left[ -(\hc^\dagger)^2
   - 2i \hc^\dagger \hd^\dagger +(\hd^\dagger)^2 \right]/2\; .
\endarr
We note that only $({\hat u}^{\prime\dagger})^2$ and $({\hat
  v}^{\prime\dagger})^2$ terms are selected by the detection of a
single photon at each detector. The output state conditioned on a
two-fold coincident count is given by
\begarr
|N,N\ra \rightarrow
 {A \over \sqrt{2}} \left[ |N,N-2\ra+ |N-2,N\ra \right]\; ,
 \label{dowling}
\endarr
where $A = \sqrt{N (N-1) /2} ~r^2 t^{2N-2}$, and the maximum
probability success is obtained when $r^2 = {1 \over N}$ and, $t^2 =
{N-1 \over N}$. 
The output state of the form Eq.\ (\ref{dowling})
shows the exact correlation for the Yurke state \cite{dowling98}.
The generation of the desired correlations described in
section \ref{yurke-sec} can thus be obtained from the dual Fock state
$|N,N\rangle$ with probability $|A|^2$.
Note that this probability has its asymptotic 
value of $1/(2e^2)$, independent of $N$.

A similar correlation can also be ontained 
by post selecting the outcome, conditioned upon
only one photon detection by either one of the two detectors.
The analysis becomes simpler than the one decribed above:
one photon either from mode $a$ or from $b$ is to be detected.
In this case, however, one needs to allow a non-detection 
protocol \cite{kok01a}
since one of the two detectors should not be fired.
 
In this section we described generation of quantum correlation
by using projective measurments,
where
the fundamental lack of which-path information
provides the entanglement between the two output modes.
Using a stack of such devices with appropriate phase shifters
has been proposed for generating
maximmally path-entangled state of the form 
Eq.\ (\ref{wineland}) \cite{kok01a}.
In atom interferometry a similar technique for the the generation 
of such a correlation has been proposed \cite{dowling98} with selective
measurements on two interfering Bose condensates \cite{castin97}.

\section{summary}

In this paper we discuss the equivalence between the optical
Mach-Zehnder interferometer, the Ramsey spectroscope, and the quantum
Fourier transform in the form of a specific quantum logic gate. Based
on this equivalence, we introduce the so-called quantum Rosetta stone. 

The interferometers and spectroscopes can yield
Heisenberg-limited phase sensitivity when suitable input states are
chosen. We presented three types of such states and showed that they
need non-local correlations of some sort. 
Path-entangled multi-photon states can be generated using projective
measurements. These states yield $1/N$ sensitivity also for small $N$,
and are essential for applications such as quantum lithography. The
dual Fock-state approach to Heisenberg-limited interferometry must be
accompanied by ``non-local'' detection schemes. The generation of a
suitable correlation from the dual Fock state via projective
measurements may be useful in that it avoids complicated signal
processing or quantum nondemolition measurements. 
One can also construct a single-photon quantum nondemolition device 
and a quantum repeater in this paradigm \cite{kok02,kok02b}.
Many more important insights are expected by the application of the
quantum Rosetta stone to quantum metrology and quantum information
processing. 

\section*{acknowledgement}

This work was carried out by the Jet Propulsion Laboratory,
California Institute of Technology, 
under a contract with the National Aeronautics
and Space Administration.
We wish to thank
D.S.\ Abrams, C.\ Adami, N.\ J.\ Cerf,
J.\ D.\ Franson, P.\ G. Kwiat, 
T.\ B.\ Pittman, Y.\ H.\ Shih, 
D.\ V. Strekalov, C.\ P.\ Williams, 
and D.\ J.\ Wineland for helpful discussions. 
We would also like to acknowledge support from the ONR,
ARDA, NSA, and DARPA.
P.K.\ and H.L.\ would like to acknowledge
the National Research Council.



\end{multicols}


\begin{thebibliography}{99}
\bibitem{note} The discussion about the Mach-Zehnder is equally
  applicable to the Fabry-Perot interferometer.
\end{thebibliography}
\end{document}